\documentclass[prl,epsfig,preprint]{revtex4}
\usepackage{epsfig}

\begin{document}

\title{Flow-induced voltage and current generation in carbon nanotubes}
\author{S. Ghosh$^{1}$, A.K. Sood$^{1}$, S. Ramaswamy$^{1}$ and N. Kumar$%
^{2} $}
\affiliation{$^{1}$Department of Physics, Indian Institute of Science, Bangalore - 560
012, India}
\affiliation{$^{2}$ Raman Research Institute, C.V. Raman Avenue,Bangalore - 560 080, India}

\begin{abstract}
New experimental results, and a plausible theoretical understanding
thereof, are presented for the flow-induced currents and voltages
observed in single-walled carbon nanotube samples. In our
experiments, the electrical response was found to be sublinear --
nearly logarithmic -- in the flow speed over a wide range, and its
direction could be controlled by an electrochemical biasing of the
nanotubes. These experimental findings are inconsistent with the
conventional idea of a streaming potential as the efficient cause.
Here we present Langevin-equation based treatment of the nanotube
charge carriers, assumed to be moving  in the fluctuating field of
ions in the flowing liquid. The resulting ``Doppler-shifted"
force-force correlation, as seen by the charge carriers drifting in
the nanotube, is shown to give a sublinear response, broadly in
agreement with experiments.
\end{abstract}

\maketitle






Single-walled carbon nanotubes (SWNT) in contact with a flowing
liquid provide a unique microfluidic system that offers a large
interfacial area of intimate atomic contact between the liquid and
the solid substrate. This can lead to a strong coupling of the
charge carriers in the nanotube to the particles in the flowing
liquid, more so if the liquid is polar or ionic in character. The
effect of this coupling is expected to be further enhanced due to
charge carrier entrainment because of the quasi-one dimensionality
of the conducting nanotubes. Recently, the flow of a variety of
liquids over SWNT bundles was studied experimentally, and was found
to generate voltage in the sample along the direction of the flow
\cite{science}. Quite unexpectedly, however, the dependence of the
voltage on the flow speed was found to be  sublinear, and could be
fitted to a logarithmic form over five decades of variation of the
speed. There has been an attempt at explaining away this
flow-induced voltage in electrokinetic terms as the streaming
potential that develops along the flow of an electrolyte through a
microporous insulator \cite{science_nanotube}. Earlier
Kr$\acute{a}$l and Shapiro \cite{kral} proposed that the shear
stress of liquid flow transfers momentum to the acoustic phonons of
the nanotube and the resulting ``phonon wind" produces a current of
carriers in the nanotube. They also suggested, qualitatively, that
the fluctuating Coulomb fields of the ions in the liquid could drag
directly the carriers in the nanotubes. To produce a flow-rate of 1
cm/s in an inter-nanotube channel of 20nm width, as demanded by the
first mechanism \cite{kral} requires an enormous pressure head,
about 10$^{13}$ dynes/cm$^{2}$. The second mechanism \cite{kral}
gives currents of order femtoAmperes. In both mechanisms
\cite{kral}, the effect is linear in the flow rate, in contrast to
the experimental findings \cite{science}. The mechanism we offer is
related to the second idea of \cite{kral} but requires neither
localization of carriers nor drag at the same speed as the ions. The
main  aim of  this paper is to report (i) our new experimental
results of the measurements of the the short-circuit electrical
current as a function of the flow-speed, (ii) ways to control the
magnitude and directionality of the flow induced current; and (iii)
a general theory for the electrical response consistent with these
new observations. Our measurements clearly rule out the
electrokinetic mechanism based on the idea of a streaming potential.
Our theory is based on a stochastic treatment of the nanotube
charge-carriers assumed to be moving under the  influence of the
correlated ionic fluctuations which are advected by the liquid flow.
It is broadly consistent with our experiments.

\begin{figure}[tbp]
\centerline{\psfig{figure=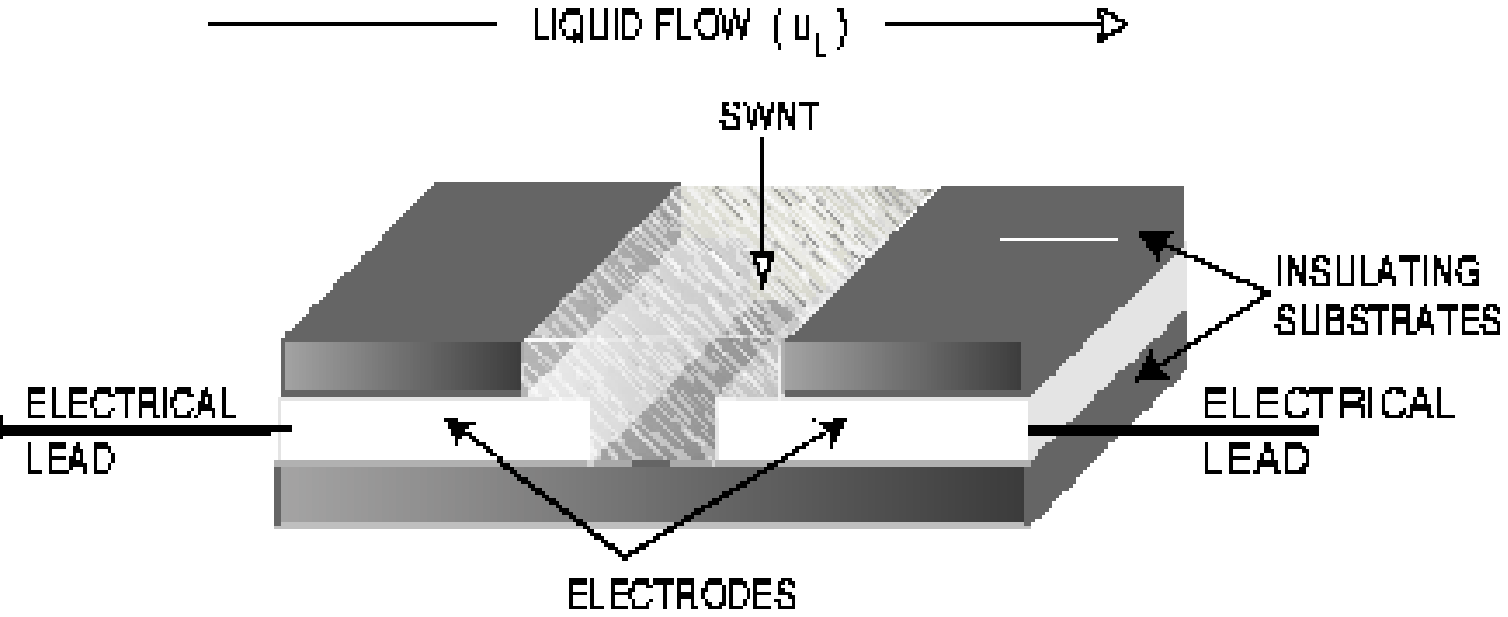,width=16cm}} \caption{Schematic
sketch of the nanotube flow sensor placed along the flow
direction($u_L$). SWNT bundles are packed between two metal
electrodes. The insulating substrate keeps the SWNT in place. The
electrical leads are taken out from the metal electrodes. }
\label{fig:sensor}
\end{figure}

\begin{figure}[tbp]
\centerline{\psfig{figure=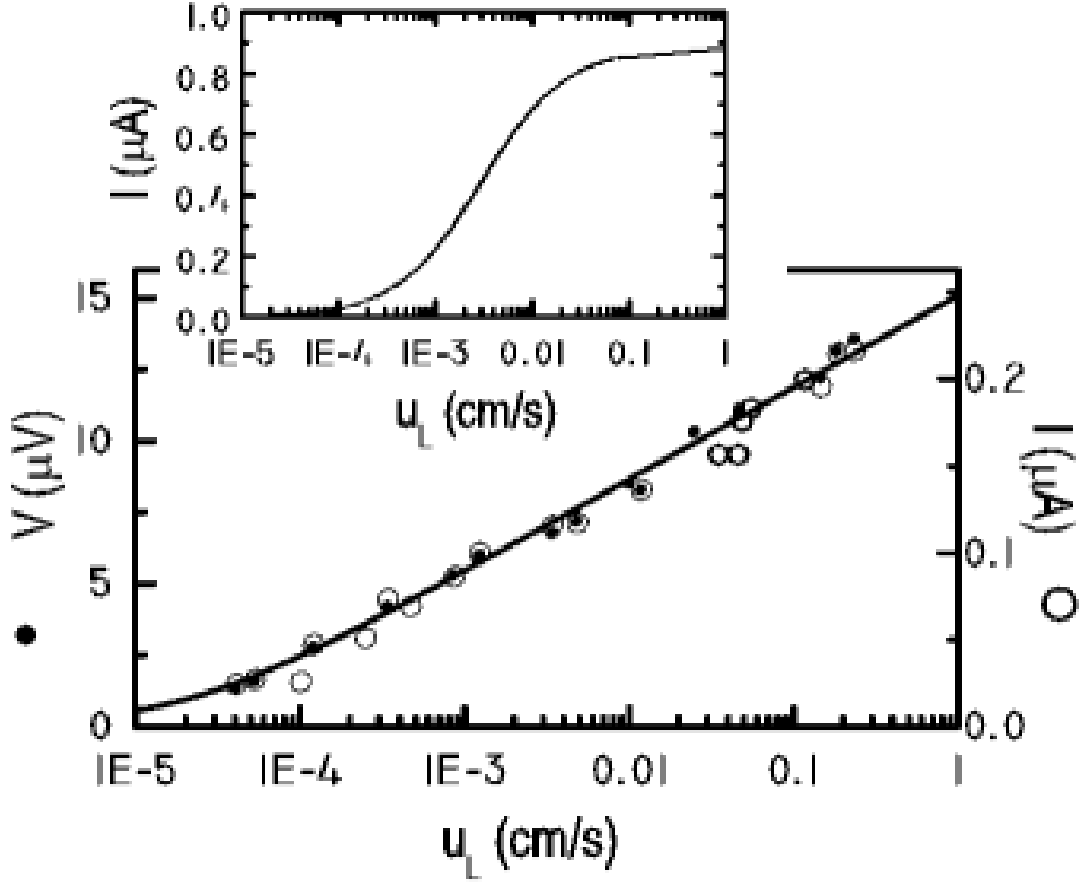,width=16cm}} \caption{Voltage
(full circles) and current (open circles) as functions of flow speed
$u_L$. The solid line is a fit to the logarithmic function as
explained in the text. Inset shows the theoretical plot of current
($I= neu_DA$) versus flow speed based on
Eqn.(\ref{mindiss},\ref{eqn:FDT1}) for
typical choice of parameters: T=300K; $\protect\epsilon$=80( CGS units); $D%
\protect\kappa=10^{-4}$ cm/s; $\protect\tau_D =10^{-16}$ s; $\protect\rho %
_{0}=10^{13}$ cm$^{-3}$; charge carrier density in nanotubes ($n = 10^{18}$
cm$^{-3}$); cross-sectional-area $(A)= 10^{-3}$ cm$^2$. The strong
sublinearity is clearly seen. }
\label{IandVvsflow}
\end{figure}

\begin{figure}[tbp]
\centerline{\epsfig{figure=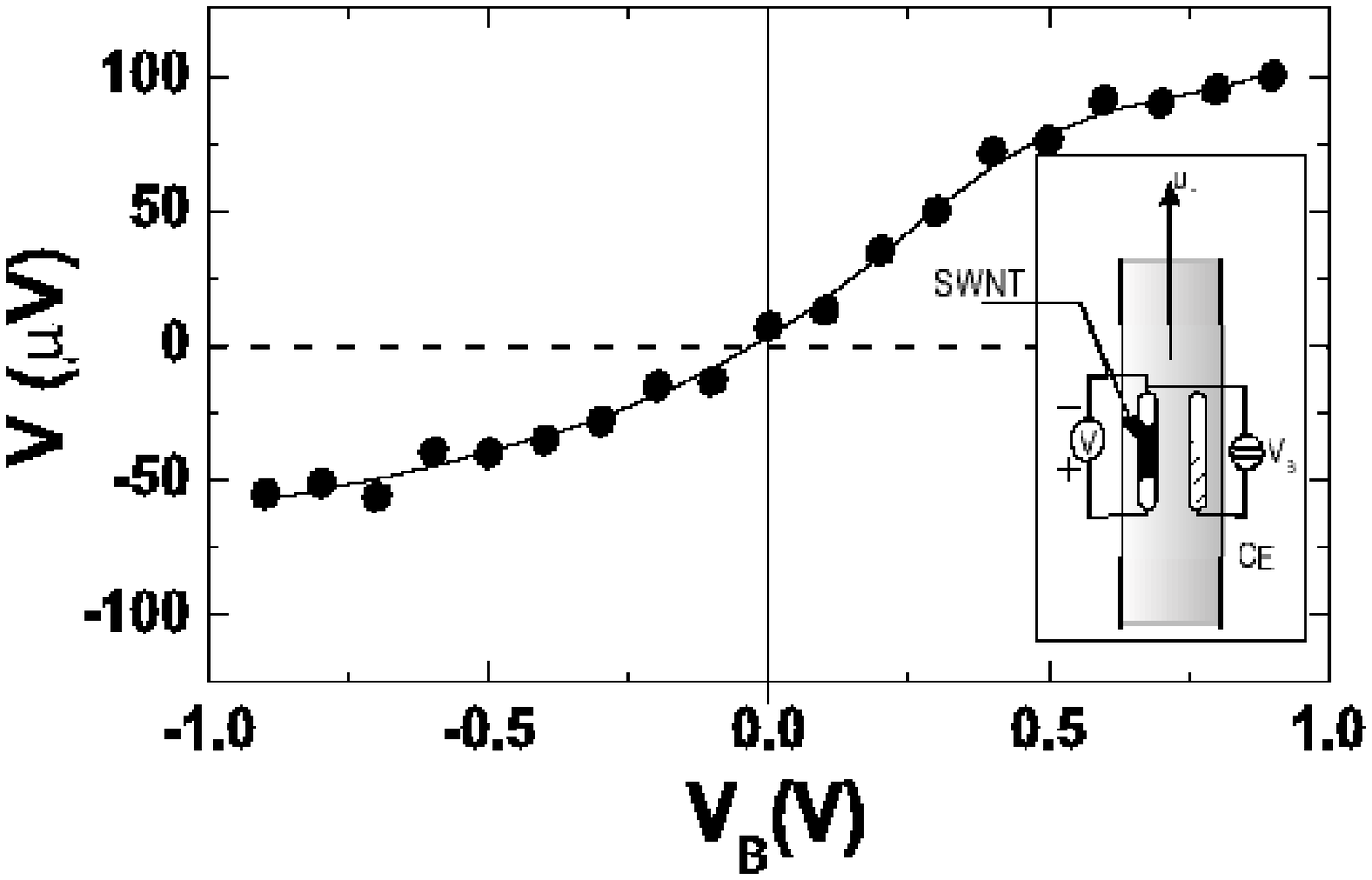,width=16cm}}
\caption{Flow-induced voltage as a function of bias $V_B$. Inset:
schematic of electrochemical biasing of the nanotubes; CE is the
counterelectrode.The solid line is a guide to the eye}
\label{biasfig}
\end{figure}

Figure \ref{fig:sensor} shows a schematic  sketch of the
construction of the flow sensor. SWNT bundles prepared by arc
discharge method \cite{pvt} are packed between two metal electrodes.
The nanotubes are kept in their place by a supporting insulating
substrate. The electrical signal is measured along the flow
direction ($u_L$) as shown Fig. \ref{fig:sensor}. The other
experimental details are as in \cite{science}. A sensor with a
minimal contact resistance  of $\sim$25$\Omega $ (found from four-
probe measurements) was used in the experiments so that the
short-circuit current could be measured. The short-circuit current
(open-circuit voltage) was measured by connecting the microammeter
(millivoltmeter) across the SWNT sample. The resistance (two-probe)
of the device, measured with the sensor dipped in the liquid was
found to be  $\sim 70$ $\Omega $. Figure~\ref{IandVvsflow} shows the
dependence of the induced voltage and current on the flow velocity
u$_{L}$. The solid line is a fit to the empirical relation $I=\alpha
_{I}\log (\beta _{I}$ $u_{L}+1),$ with $\alpha _{I}=0.02\mu A$ and
$\beta
_{I}=4.8\times 10^{4}s/cm.$ The voltage also fits the empirical relation $%
V=\alpha _{V}\log (\beta _{V}$ $u_{L}+1),$ where $\alpha _{V}=1.4\mu V$ and $%
\beta _{V}=4.8\times 10^{4}s/cm.$ It can readily be seen that
$\alpha _{V}=\alpha _{I}\times R$, i.e., the resistance encountered
is precisely the  resistance (2 probe) of the device. This is an
important point to note: if an electrokinetic mechanism were
operating, the resistance obtained would have been orders of
magnitude higher, i.e., equal to that of the electrolyte ($\sim
0.1M\Omega $)\cite{electrochembook}. This in itself rules out quite
decisively the electrokinetic mechanism of voltage generation. Next,
we consider the measured direction of the flow induced current with
respect to the flow direction as a function of the bias voltage
$V_{B}$ (see inset of Fig.\ref{biasfig}). This potential biases the
SWNT with respect to the Au-reference electrode immersed in the flow
chamber close to sample as shown in the inset of
Fig.(\ref{biasfig}). The dependence of the sign and the magnitude of
the flow-induced voltage on $V_{B}$ for an aqueous solution of
$0.01$ M KCl (conductivity $1.4$ mS/m) and for a fixed flow speed of
0.04 cm/s is shown in Fig.(\ref{biasfig}). It is seen that the
flow-induced signal is positive, i.e., $I$ is anti-parallel to
$u_{L}$ when $V_{B}$ is positive, and the sign of the signal is
reversed, i.e., $I$ is parallel to $u_{L}$, for $V_{B}$ negative,
consistent with our theory. Further, in our theoretical mechanism,
based on fluctuations, the carriers in the nanotube drift
necessarily in the same direction as the flow velocity $u_{L}$,
quite independently of the sign of the ionic charge. Thus, for the
current $I$ to be parallel (antiparallel) to $u_{L}$ the charge
carriers in the nanotubes need to be holes (electrons). When the
nanotubes are biased positively, the anions (Cl$^{-}$, OH$^{-}$)
move closer to the SWNT, localizing holes on the carbon and making
electrons available for flow-induced current. Similarly, holes are
liberated when the bias is negative \cite{Electrochemical.APL}. As
the bias voltage is increased the number of carriers participating
in the flow-induced current will increase as shown in
Fig.(\ref{biasfig}). Thus, the results obtained by electrochemically
biasing the nanotubes are naturally consistent with our mechanism of
(coulombic) forcing of the nanotube charge carriers by the liquid
flow. The dependence of the flow-induced signal on the concentration
of different types of ions in the liquid is, however, found to be
complicated and non-monotonic. These details will not be addressed
here.

We now turn to our theoretical mechanism for the flow-induced
current/voltage. This can be understood qualitatively in terms of three
physically distinct but related ideas: (a) \emph{Induced friction}: The
fluctuating charge density of the ions close to the nanotube couples
couloumbically to the charge carriers in the nanotube and, therefore, offers
a friction to the motion of these charge carriers (in addition to the Ohmic
friction intrinsic to the carbon nanotubes). This, of course, follows
directly from the fluctuation-dissipation theorem; (b) \emph{%
Flow-induced drag}: In virtue of the above frictional coupling, an imposed
liquid flow drags the charge carriers along through the nanotube; (c) \emph{%
Reduction of induced friction at high flow speeds}: The space-time
correlated couloumbic fluctuations, inherent to the liquid electrolyte, are
\emph{advected} by the liquid flow, and thus get Galilean boosted (Doppler
shifted) as seen in the mean rest frame of the drifting carriers in the
nanotube. Correspondingly, as we will see, the friction they offer to the
motion of the charge carriers in the nanotube diminishes with increasing
flow speed. This is crucial to the observed sublinear dependence of the
charge drift-velocity (electrical response) on the liquid flow speed.With
the above in mind, we will now derive these frictional effects, first from a
heuristic argument, and then analytically from a Langevin-equation treatment.

Consider a nanotube placed along the the $z$-axis with the liquid flowing
parallel to it. We model the nanotube as a classical one- dimensional (1D)
conductor with diffusive charge transport. In the steady state, let $u_{L}$
and $u_{D}$ be, respectively, the velocity of liquid(L) flow and the induced
drift velocity of charge carriers in the nanotubes, all measured relative to
the nanotube lattice. Now, in the absence of any frictional coupling to the
liquid flow, the drift velocity $u_{D}$ will tend to relax to zero, i.e., to
rest with respect to the lattice, with a relaxation time $\tau _{D}$ \
characteristic of the nanotube resistivity. Similarly, if we were to
``switch off'' the resistive coupling to the lattice, the drift velocity $%
u_{D}$ would relax to the liquid velocity $u_{L}$, i.e., to the rest frame
co-moving with the liquid, because of the frictional coupling (drag), with a
relaxation time $\tau _{L}$. Now, therefore, in the presence of both these
frictional influences, the drift velocity $u_{D}$ will assume a steady-state
value $u_{D}<u_{L}$ that satisfies $u_{D}/\tau _{D}=(u_{L}-u_{D})/\tau _{L}$
giving
\begin{equation}
u_{D}=u_{L}/(1+\tau _{L}/\tau _{D}).  \label{mindiss}
\end{equation}
Equation (\ref{mindiss}) is merely a restatement of the condition of
frictional force-balance in the steady state. It would appear to give an
induced short-circuit current (equivalently, an open circuit voltage via the
nanotube resistance) along the nanotube which is linear in the flow
velocity. The nonlinearity is, however, really hidden in the $u_{L}$
dependence of the relaxation time $\tau _{L}$ that we will now try to make
explicit. It may be noted here that we are assuming, for simplicity, a
uniform liquid flow without the hydrodynamic complications of a no-slip
boundary condition.

In a simple caricature of the real situation then, consider the ionic
density in the liquid, fluctuating thermally and flowing past the nanotube
at a mean velocity $u_{L}\hat{\mathbf{z}}$, producing thereby a fluctuating
couloumbic potential $\phi (\mathbf{r},t)$, at a point $\mathbf{r}$ at time $%
t$. We are, of course, interested in the case of $\mathbf{r}$ lying on the $%
z $ axis i.e., $\mathbf{r}=(0,0,z)$ (in the 1D nanotube). For the space-time
correlation function $\langle \phi (0,0)\phi (\mathbf{r},t)\rangle \equiv
G_{0}(\mathbf{r},t)$ in the mean rest frame of the ions, the charge carriers
in the nanotube see this correlation Galilean boosted to $G(\mathbf{r}%
,t)\equiv G_{0}(\mathbf{r}-\hat{\mathbf{z}}vt,t)$ with
$v=u_{L}-u_{D}$. This Galilean boost (Doppler shift) is the key
physical point of our treatment. At $u_L = 0$, the
fluctuation-dissipation theorem (FDT)   tells us that the
coefficient of the zero-frequency friction to the motion of the
charge carriers in the nanotubes, arising from the ionic thermal
fluctuations, is proportional to the time integral of this on-site
force-force correlation function. If we \emph{assume} this relation
even for $u_L \neq 0$ we have
\begin{equation}
1/\tau _{L}=1/(m_{e}k_{B}T)\int_{-\infty }^{\infty }\langle eE_{z}(\mathbf{r}
-\hat{\mathbf{z}}vt,t) eE_{z}(r,0)\rangle dt.  \label{eqn:FDT}
\end{equation}%
%
Here $E_{z}$ is the $z$-component of the coulombic (electric ) field due to
the ions; $m_{e}$ is the mass of the charge carrier with $e$ the electronic
charge; $k_{B}$ is the Boltzman constant, and $T$ the absolute temperature.
We re-write the right hand side of Eqn.(\ref{eqn:FDT}) in Fourier ($\mathbf{q%
}$)- space, expressing the above force-force correlator in terms of the
ionic charge-densities $\rho(\mathbf{r},t)$ using $E_{z}(\mathbf{q}%
,t)=-iq_{z}\phi (\mathbf{q},t)$~and~$-q^{2}\phi (\mathbf{q})=e\rho (\mathbf{q%
},t)/\epsilon$, where $\epsilon$ is the solvent dielectric constant, and
obtain straightforwardly 
\begin{eqnarray}
\frac{1}{\tau _{L}} &=& \frac{1}{m_{e}k_{B}T} \left( \frac{4\pi e^{2}}{%
\epsilon }\right) ^{2}\rho _{0}\int \frac{d\mathbf{q}}{\left( 2\pi \right)
^{3}}\left( \frac{q_{z}^{2}}{q^{4}}\right) \left( \frac{q^{2}}{q^{2}+\kappa
^{2}}\right) \left( \frac{2/\tau _{q}}{\left( vq_{z}\right) ^{2}+\left(
1/\tau _{q}\right) ^{2}}\right)  \label{eqn:FDT0} \\
&=&\frac{\rho _{0}}{4\pi m_{e}D\kappa k_{B}T} \left( \frac{4\pi e^{2}}{%
\epsilon }\right) ^{2} \frac{1}{x}\left( 1-\frac{2}{x}+\frac{1}{x^{2}}\log
\left| 1-x^{2}\right| -\frac{1}{x^{2}}\log \left| \frac{1-x}{1+x}\right|
\right)  \label{eqn:FDT1}
\end{eqnarray}%
%

where $x=v/D\kappa $, and $\rho_0$ is the mean ionic number density. In
Eqns. ({\ref{eqn:FDT0}) and (\ref{eqn:FDT1}}) we have used the Debye
-screened form for the static charge structure factor $S_{q}^{_{^{0}}}=%
\langle \rho_{q} \rho_{-q} \rangle= q^{2}/(q^{2}+\kappa ^{2})$ with
screening length $\kappa^{-1} $ as the inverse of the Debye
screening length and a diffusive form $1/\tau _{q}=Dq^{2}$
\cite{note1} with $D$ the ionic diffusion constant. It can be seen
at once that Eqn.(\ref{eqn:FDT1}) , taken in conjunction with
Eqn.(\ref{eqn:FDT}), gives a drift velocity $u_{D}$(and therefore
the short-circuit current) as a  sublinear function of the flow
velocity $u_{L}$. This sublinearity is a generic feature of this
mechanism, and is clearly seen in inset of Fig.(\ref{IandVvsflow})
in which we have plotted the induced current versus the liquid flow
speed derived directly from Eqn.(\ref{eqn:FDT1}) for certain choice
of parameters. Rather than attempting to justify this use of the
regression formula in a nonequilibrium context, we turn instead to a
direct evaluation of the drift speed starting from a Langevin
equation. We recover the same sublinear functional form for the
response as in Eqn.(\ref{eqn:FDT0}).

The one-dimensional (1D) position $z(t)$ of a tagged charge carrier in the
nanotube obeys the overdamped Langevin equation
\begin{equation}
\zeta {\frac{dz}{dt}}=F(z(t),t)+f(t)  \label{langeq}
\end{equation}%
%
%
%
where $F(z(t),t)$ is the fluctuating force the ions exert on the carrier at $%
z(t)$ in the nanotube, and $f$ and $\zeta $ are the thermal Gaussian white
noise and friction intrinsic to the nanotube, with $\langle f(0)f(t)\rangle
=2k_{B}T\zeta \delta (t)$. Strictly speaking, $F(z(t),t)$ should be a
function of the instantaneous separations between the positions $z(t)$ of
the tagged carriers in the nanotube and those of all the ions in the ambient
liquid. We would like to replace this complex many-body problem by an
effective Langevin equation for the dissipative motion of the charge
carriers in the nanotube, without invoking the FDT which, strictly speaking,
holds only in the absence of the flow. Our treatment nonetheless captures
the qualitative physics of the dissipative entrainment of carriers by the
moving ions. For this, we simply take $F(z,t)$ to be a Gaussian random force
field, with zero mean and a two-point correlation $C(z,t)$. Then from Eqn.(%
\ref{langeq}), the drift velocity of the carrier $u_{D}\equiv \langle {%
dz(t)/dt} \rangle =(1/\zeta )\langle F(z(t),t)\rangle $ can be rewritten,
using Novikov's theorem \cite{gardiner} in relation to the Gaussian noise $%
F(z,t)$, as
\begin{eqnarray}
u_{D} &=&{\frac{1}{\zeta }}\int_{z,z^{\prime },t^{\prime }}C(z-z^{\prime
},t-t^{\prime })\langle {\frac{\delta \,\delta (z-z(t))}{\delta F(z^{\prime
},t^{\prime })}}\rangle  \nonumber \\
&=&-{\frac{1}{\zeta }}\int_{z,z^{\prime },t^{\prime }}C(z-z^{\prime
},t-t^{\prime })\langle {\frac{\delta z(t)}{\delta F(z^{\prime },t^{\prime })%
}}\delta ^{\prime }(z-z(t))\rangle .  \label{novik}
\end{eqnarray}
Using Eqn.(\ref{langeq}) to evaluate the functional derivative in Eqn.(\ref%
{novik}), and writing $C(z,t)$ in terms of its Fourier transform $C_{q}(t)$
yields
\begin{eqnarray}
u_{D} &=&{\frac{1}{\zeta ^{2}}}\int_{-\Lambda }^{\Lambda }{\frac{\mbox{d}%
\mathbf{q}}{(2\pi )^{3}}}\int_{0}^{\infty }\mbox{d}t\langle
e^{iq_{z}[z(t)-z(t^{\prime })]}\rangle iq_{z}C_{q}(t) \   \label{driftgenq}
\end{eqnarray}%
%
%
%

where 
$\Lambda $ is an ultraviolet cut-off of the order of an inverse ionic
diameter. As in the preceding heuristic treatment, let us take the
correlation $C_{q}(t)$ of the ions to be the Galilean boost, with velocity $%
u_{L}$, of an equilibrium correlation function $C_{q}^{0}(t)$, with a
relaxation time $\tau _{q}$:
\begin{equation}
C_{q}(t)=C_{q}^{0}(t)e^{-iq_{z}u_{L}t}\equiv
C_{q}^{0}e^{-iq_{z}u_{L}t}e^{-t/\tau _{q}},  \label{cqt}
\end{equation}%
%
%
%
where $C_{q}^{0}$ is the equilibrium equal-time correlation function of the
force fluctuations. This form, despite its undeniable limitations, is the
simplest way to capture the basic physics of ions moving past the nanotube,
and admits an essentially analytical treatment. As before, the force-force
correlation $C_{q}^{0}$ can be expressed in terms of the ionic
charge-density correlation, which is known as an input from the liquid
state(dilute ionic solution) theory, namely that $C_{q}^{0}\propto
(q_{z}^{2}/q^{4})S_{q}^{0}$ with the ionic charge structure factor $%
S_{q}^{0}=q^{2}/(q^{2}+\kappa ^{2})$. Note the factor $(q_{z}^{2}/q^{4})$
arising from the gradient ($\partial/\partial z$) and the Laplacian($\nabla
^2$) in $q$- space. Replacing $z(t)$ in Eqn.(\ref{driftgenq}) by its mean $%
u_Dt$ for simplicity, we obtain the compact expression
\begin{equation}
u_{D}=v\alpha\int_{-\Lambda }^{\Lambda }{\frac{\mbox{d}\mathbf{q}}{(2\pi
)^{3}}}\left( {\frac{1}{q_{z}^{2}v^{2}+\tau _{q}^{-2}}}\right) \left( {\frac{%
q_{z}^{2}}{q^{2}+\kappa ^{2}}}\right),  \label{driftvelspecial}
\end{equation}%
%
%
for the drift velocity of the charge carriers in the nanotube, where as
before $v=u_{L}-u_{D}$, $\alpha$ is a lumped constant of proportionality
that depends on the parameters of the liquid-state correlation function
input used above. With the ultraviolet cut-off ($\Lambda$) set to infinity,
and \cite{note1} with $1/\tau_q = Dq^2$, Eqn. (\ref{driftvelspecial}) has
precisely the form of Eqn.(\ref{mindiss}) taken in conjunction with the Eqn.(%
\ref{eqn:FDT1}), allowing us thereby to identify the integral on its
right-hand side essentially with $\tau _{D}/\tau _{L}$. This gives us an
expression for the flow-speed dependence of $\tau _{L}$, and thus finally an
analytic expression for the charge drift velocity ($u_D$) as a function of
the liquid flow velocity ($u_L$). This reaffirms our heuristic argument
given at the beginning.

We close by summarizing the main points of our work. First, on the
experimental side, we have clearly shown that the liquid flow
produces not only a voltage (i.e., not merely a capacitive
charging), but a short-circuit current as well in the nanotube; that
both have a  sublinear dependence on the imposed flow-speed; and
that the voltage/current ratio corresponds to the nanotube sample
resistance. These observations are incompatible with an
electrokinetic origin for the (electronic) current in the nanotube.
On the theoretical side, we have proposed a theory wherein the
current is essentially a statistical consequence of the flow-induced
asymmetry in the correlation of the ions, in the ambient fluid as
seen by the charge carriers in the nanotube. Importantly, our theory
predicts in general a  sublinear behavior for the electrical
response, with a linear regime at only the smallest values of
imposed flow. The extended logarithmic regime seen in experiments
can presumably be rationalized in detail with particular forms for
the correlation function ($C_q^0$) and the relaxation time
($\tau_q$), as inputs to be taken from the liquid state (dilute
ionic solution) theory. Moreover, a realistic treatment will require
taking into account details of the complex, hydrophobic,
inter-nanotube micro-fluidic environment of our mat samples. Thus,
very specifically, the no-slip boundary condition would imply a
decreasing velocity of the flow nearer the nanotube (the shear
flow).This decrease in the flow velocity will,however,be offset by
the corresponding increase in its effectiveness(via the screened
coulombic forcing) closer to the interface.The resulting levelling
is expected to broaden the sublinear response and thus improve
agreement with the experiment. Our main point, however, is that the
experiments show sublinear behavior, which is inconsistent within
existing theories of flow-induced voltages and currents in
nanotubes, and that our approach naturally and inevitably leads to
strong sublinearity. Finally, we emphasize that the flow-induced
asymmetry of the random fluctuations is key to the charge-carrier
drift (drag) mechanism in our theory. In this broad sense our
approach here subsumes the asymmetric fluctuating ratchets invoked earlier %
\cite{science} in a general way. In this connection, reference must
be made to the idea of a drag (shear) induced by the relative motion
between material surfaces, where the Doppler shifted and aberrated
photonic fluctuations, e.g. zero-point photons, have been invoked
very effectively \cite{pendry,2D:electron:gas} . We would also like
here to add that more than one mechanisms could very well be at work
in these systems. For example, in a recent publication Persson et al
\cite{persson} have invoked  a combination of frictional stick-slip
and barrier-hopping to explain the observed phenomenon of
flow-induced voltages in SWNT. The one we propose here seems
particularly robust and general, and we look forward to experimental
tests, especially of the predicted saturation of the electrical
response at high flow speeds.

SR (through the Center for Condensed Matter Theory) and AKS thank the DST,
India, for support. AKS thanks Prof. C.N.R. Rao for nanotube samples and
fruitful collaboration on nanotubes.

\newpage

\newpage

\begin{thebibliography}{99}
\bibitem{note1} Consistency demands $\tau_q = S_q^0/Dq^2$; we have checked
numerically that this form also produces qualitatively the same shape as in
the inset to Fig.(1).
\end{thebibliography}
\end{document}